\documentclass[print, twocolumn,showpacs]{revtex4}
\usepackage{amsmath}
\usepackage{txfonts}
\usepackage{graphicx}
\usepackage{bm}
\usepackage{epsfig}
\usepackage{setspace}
\begin{document}
\draft
\bibliographystyle{revtex4}

\titlepage
\title{Decoherence from a spin-chain with three-site interaction}
\author{Li-Jun Tian\footnote{Email: tianlijun@staff.shu.edu.cn.},Cai-Yun Zhang,Li-Guo Qin}
\affiliation {Department of Physics, Shanghai University, Shanghai,
200444, China}

\begin{abstract}
We investigate the time evolution of quantum discord and entanglement for two-qubit coupled to a spin chain with three-site interaction in the weak-coupling region. If the quantum system evolves from a Bell state, quantum correlations decay to zero in a very short time at the critical point
of the environment. We found there exist some special interval of the three-site coupling strength in which the decay of quantum discord and entanglement can be delayed. When the qubits are initially prepared in a Bell diagonal state, the decay of entanglement is also delayed in the special interval, but the decay of quantum discord is enhanced. Besides, the sudden transition between classical and quantum decoherence is observed. we found the transition time can be lengthened at the special range of three-site interaction and shorten by degree of anisotropy.

\end{abstract}

\maketitle

\section{\label{sec:level1}INTRODUCTION}
Entanglement, which is an important characteristic trait of quantum mechanics, is playing a key role in quantum information processing \cite{ES}. It is a type of quantum correlation arising from the superposition principle of quantum mechanics and has a wide range of applications on quantum teleportation \cite{CHB}, quantum cryptography \cite{NG}, and universal quantum computing \cite{DPD}. In fact, it is only one aspect of quantum correlation and there exist other nonclassical correlations \cite{HOWHZ,SL,BDVVB,KMTPWSVVMW,ASHKDB}. Recent studies have indicated non-entanglement quantum correlation can be responsible for the quantum computational speed of deterministic quantum computation with one qubit \cite{ADASCMC}. Quantum discord was introduced by Olliver and Zurek to quantify the nonclassical correlation of bipartite system and it is defined as the difference between two quantum versions of classical mutual information \cite{HOWHZ}.

Quantum discord has been studied widely, such as its dynamical property and indicating quantum phase transition \cite{TCGG,ADASCMC,BPL,TWSSFFFCJVB,LCWJSXXY,HJFJGLJZBS,LJTYYYLGQ}. It has been shown thermal quantum discord in Heisenberg models is different from thermal entanglement in many unexpected ways, and the results point out that it may be useful in the experimental detection of critical points for quantum phase transition \cite{TCGG}. Quantum discord could be considered as a figure of merit for the quantum advantage in some computational models with little or zero entanglement \cite{ADASCMC,BPL}. T. Werlang has pointed out that quantum discord is more robust than entanglement against decoherence and concluded quantum algorithms based on quantum discord can be more robust than those based on entanglement \cite{TWSSFFFCJVB}. Ref. \cite{LCWJSXXY} show that quantum discord may reveal more general information than quantum entanglement for characterizing the environment's quantum phase transition.

For a realiatic quantum syatem, it is inevitable to interact with the environment and the interactions between the quantum system and its environment leads to decoherence which leads to the destruction of quantum properties. The decoherence behaviors of correlation have been investigated in Markovian and non-Markovian environment and the results show that quantum discord vanishes only in the asymptotic limit in all cases under the Markovian environments \cite{TWSSFFFCJVB} and instantaneously disappear at some time points under a non-Markovian environment \cite{FFFTWCABLGEAAOC}. Moreover, it is found either the classical or the quantum part of correlation may be unaffected by decoherence \cite{AFLADCFMCAA,BWZYXZQCMF}. In addition, the dynamic behavior of quantum
discord under spin environment has been discussed \cite{BQLBSJZ,QLWAM}. It is found that the quantum discord for the two central
qubits can become minimized rapidly close to the critical point of a quantum phase transition \cite{BQLBSJZ}. Besides, the DM interaction is considered in spin environment. The results imply the DM interaction can enhance the decay of quantum correlations and this effect can be strengthened by anisotropy parameter \cite{QLWAM}. Recently, studies of decoherence induced by spin environment have received a great deal of attention \cite{QHTSZLXFZPSCP,YZGZPLSS,CWWLJM} and the results revealed that quantum coherence is dramatically destroyed at the critical point of the quantum phase transtion of the external environment \cite{YZGZPLSS}.

Previous studies have been shown that the three-site interaction may play an important role in quantum entanglement and discord of the bipartite system \cite{JLGHSS,WWCJML,BQLBSJZC}. Thus, it is necessary to consider when environment presents three-site interaction, how it affects the evolution of quantum discord of the central two-qubit system. In this paper, we offer a comparative study of the dynamic behaviors of quantum discord and entanglement for two-qubit prepared in a class of X-structure state under spin chain environment. We will show how the three-site interaction affects those process of the nonclassical correlation. The paper is organized as follows. We start in section II by introducing and diagonalizing the model, and obtain the reduced density matrix of the central two-qubit system. In section III, we analytically and numerically evaluate quantum discord and entanglement, then present the main results. The last section is devoted to the conclusions.

\section{\label{sec:level2} MODEL AND ITS SOLUTION}

The total Hamiltonian for two central qubits coupled to an XY spin chain with three-site interaction is composed of two parts: the interaction Hamiltonian $H_{I}$ of the central two-qubit with the chain and the self-Hamiltonian $H_{E}$ of the spin-chain environment. They can be written as
\begin{eqnarray}
H&=&H_{I}+H_{E},\nonumber\\
H_{E}&=&-\sum_{l}^{N}(\frac{1+\gamma}{2}\sigma_{l}^{x}\sigma_{l+1}^{x}+\frac{1-\gamma}{2}\sigma_{l}^{y}\sigma_{l+1}^{y}+\lambda\sigma_{l}^{z})\nonumber\\
&&-\sum_{l}^{N}\alpha(\sigma_{l-1}^{x}\sigma_{l}^{z}\sigma_{l+1}^{y}+\sigma_{l-1}^{y}\sigma_{l}^{z}\sigma_{l+1}^{x}),\nonumber\\
H_{I}&=&-\sum_{l}^{N}{g}(\frac{1+\delta}{2}\sigma_{A}^{z}+\frac{1-\delta}{2}\sigma_{B}^{z})\sigma_{l}^{z},
\end{eqnarray}
where $g\left(\frac{1\pm\delta}{2}\right)$ denotes coupling strength. $\sigma_{A(B)}^{z}$ and $\sigma_{l}^{a} (a=x,y,z)$ are Pauli matrices of the two central qubits and the surrounding chain, respectively. $N$ is the total number of the sites of the chain. The parameter $\lambda$ characterizes the strength of the transverse field, $\gamma$ defines the degree of anisotropy of the interactions in the $x-y$ plane, and $\alpha$ denotes the strength of three-site interaction.

The eigenstates of the operator $(\frac{1+\delta}{2}\sigma_{A}^{z}+\frac{1-\delta}{2}\sigma_{B}^{z})$ are simply given by
\begin{equation}
|\phi_{1}\rangle=|00\rangle,|\phi_{2}\rangle=|01\rangle,|\phi_{3}\rangle=|10\rangle,|\phi_{4}\rangle=|11\rangle,
\end{equation}
and the Eq. (1) can be written as
\begin{equation}
H=\sum_{\mu=1}^{4}|\phi_{\mu}\rangle\langle\phi_{\mu}|\otimes H_{E}^{\lambda_{\mu}}.
\end{equation}
where the parameter $\lambda_{\mu}$ is $\lambda_{{1}(4)}=\lambda\pm g, \lambda_{{2}(3)}=\lambda\pm g\delta,
$ and $H_{E}^{\lambda_{\mu}}$ is obtained from $H_{E}^{\lambda}$ by replacing $\lambda$ with $\lambda_{\mu}$.
The central two-qubit system and the chain environment are supposed to be initially uncorrelated with the density operator $\rho^{\mathrm{tot}}(0)=\rho_{AB}(0)\otimes|\varphi\rangle_{EE}\langle\varphi|$, where $\rho_{AB}(0)$ and $|\varphi\rangle_{E}$ are the initial states of the two central qubits and the spin chain environment, respectively. The time evolution of the total system is governed by $\rho^{\mathrm{tot}}(t)=U(t)\rho^{\mathrm{tot}}(0)U^{\dagger}(t)$ with $U\left(t\right)=\exp\left(-iHt\right)$. To obtain analytical expression of the time evolution operator, we need to diagonalize the projected Hamiltonian $H_{E}^{\lambda_{\mu}}$. With the help of Jordan-Wigner transformation given by
\begin{eqnarray}
\sigma_{l}^{x} &=& \prod_{s<l}(1-2c_s^\dag c_s)(c_l+c_l^\dag), \nonumber\\
\sigma_{l}^{y} &=& -i\prod_{s<l}(1-2c_s^\dag c_s)(c_l-c_l^\dag), \nonumber\\
\sigma_{l}^{z} &=& 1-2c_l^\dag c_l, \nonumber
\end{eqnarray}
where $c_l$ and $c_l^\dag$ are spinless fermion annihilation and creation operator respectively and employing Fourier transforms of the fermionic operators described by $d_{k}=\frac{1}{\sqrt{N}}\Sigma_{l}c_{l}e^{-i2\pi lk/N}$ with $k=-M\cdots M$ and $M=(N-1)/2$, following by Bogoliubov transformation $\gamma_{k,\lambda_{\mu}}=\cos\frac{\theta_{k}^{\lambda_{\mu}}}{2}d_{k}-i\sin\frac{\theta_{k}^{\lambda_{\mu}}}{2}d_{-k}^{\dagger}$ with $\theta_{k}^{\lambda_{\mu}}=\arctan\left\{ \gamma \sin\left(\frac{2\pi k}{N}\right)/\left[\lambda_{\mu}-\cos\left(\frac{2\pi k}{N}\right)\right]\right\} $, we obtain the diagonalized Hamiltonian as
\begin{equation}
H_{E}^{\lambda_{\mu}}=\Sigma_{k}\Lambda_{k}^{\lambda_{\mu}}\left(\gamma_{k,\lambda_{\mu}}^{\dagger}\gamma_{k,\lambda_{\mu}}-\frac{1}{2}\right),
\end{equation}
where the spectrum reads
\begin{equation}
\Lambda_{k}^{\lambda_{\mu}}=2\left[\varepsilon_{k}^{\lambda_{\mu}}+\alpha \sin\left(\frac{4\pi k}{N}\right)\right],
\end{equation}
in which $\varepsilon_{k}^{\lambda_{\mu}}=\left\{ \left[\cos\left(\frac{2\pi k}{N}\right)-\lambda_{\mu}\right]^{2}+\gamma^{2}\sin^{2}\left(\frac{2\pi k}{N}\right)\right\} ^{\frac{1}{2}}$

With the calculation above we can obtain the time evolution operator for the Hamiltonian:
\begin{equation}
U\left(t\right)=\sum_{\mu=1}^{4}|\phi_{\mu}\rangle\langle\phi_{\mu}|\otimes U_{E}^{\lambda_{\mu}}\left(t\right),
\end{equation}
where $U_{E}^{\lambda_{\mu}}\left(t\right)=\exp\left(-iH_{E}^{\lambda_{\mu}}t\right)$ denotes the projected time evolution operator. Then the reduced density matrix of two central qubits is obtained by tracing out the environment,
\begin{eqnarray}
\rho_{AB}(t)&=&\mathrm{Tr}_{E}[\rho^{tot}(t)]\nonumber\\
&=&\sum_{\mu,\nu}^{4}F_{\mu\nu}(t)\langle\phi_{\mu}|\rho_{AB}(0)|\phi_{\nu}\rangle|\phi_{\mu}\rangle\langle\phi_{\nu}|,
\end{eqnarray}
where $F_{\mu\nu}\left(t\right)=\langle\varphi_{E}|U_{E}^{\dagger\lambda_{\nu}}\left(t\right)U_{E}^{\lambda_{\mu}}\left(t\right)|\varphi_{E}\rangle$. Now we assume that the two central qubits are initially prepared in an X-structure state
\begin{equation}
\rho_{AB}\left(0\right)=\frac{1}{4}\left(I+\sum_{m}c_{m}\sigma_{A}^{m}\otimes\sigma_{B}^{m}\right),
\end{equation}
where $m=1,2,3$ and $I$ is the identity operator. In order to expediently obtain an analytical result, we set the parameters $c_m$ to be real. According to Eqs. (7) and (8), in the basis $\left\{ |00\rangle,|01\rangle,|10\rangle,|11\rangle\right\}$, the reduced density matrix of the two central qubits is given by
\begin{eqnarray}
\rho_{AB}(t)&=&\frac{1}{4}\begin{pmatrix}
1+c_3 &0 &0 &G\\
0 &1-c_3 &W &0\\
0 &W^*_3 &1-c_3 &0\\
G^* &0 &0 &1+c_3,
\end{pmatrix}
\end{eqnarray}
where $G=(c_1-c_2)F_{14}(t)$, $W=(c_1+c_2)F_{23}(t)$, $F_{\mu\nu}(t)$ is the decoherence factor, and $\ast$ denotes complex conjugation. It is necessary  to obtain the exact expression for the decoherence factor. Then the groundstates of pure spin-chain Hamiltonian $H_E^\lambda$ and the qubit-dressed Hamiltonian $H_E^{\lambda_\mu}$  can be denoted with ${|G \rangle}_\lambda$ and ${|G \rangle}_{\lambda_\mu}$. We assume that the initial state of the environment $|\phi\rangle_{E}$ is the ground state ${|G \rangle}_\lambda$ of the pure spin-chain Hamiltonian, ${|G \rangle}_\lambda=\otimes_{k=1}^M(\cos\frac{\theta_k^\lambda}{2}){|0\rangle}_k{|0 \rangle}_{-k}+i\sin\frac{\theta_k^\lambda}{2}){|1\rangle}_k{|1 \rangle}_{-k}.$  By using the transformation ${|G \rangle}_\lambda=\prod_{k=1}^M(\cos{\Theta_k^{\lambda_\mu}}+i\sin{\Theta_k^{\lambda_\mu}} \gamma^{\dagger}_{k,\lambda_\mu}\gamma^{\dagger}_{-k,\lambda_\mu}){|G \rangle}_{\lambda_\mu}$ with $\Theta_k^{\lambda_\mu}=(\theta_k^{\lambda_\mu}-\theta_k^\lambda)/2$, We can obtain the decoherence factor \cite{JLGHSS},

\begin{figure}
  \includegraphics[width=10cm]{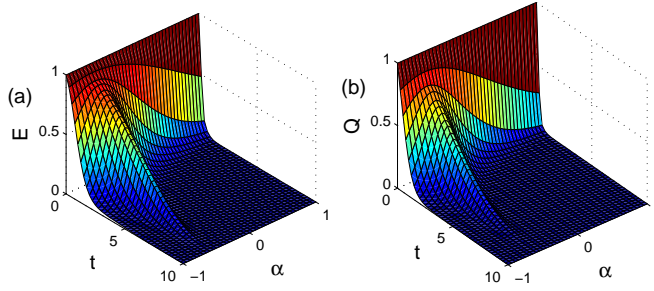}\\
\caption{(a) Entanglement and (b) quantum discord as a function of time $t$ and three-site interaction $\alpha$, Other parameters
are set as $\gamma$= 1, $\delta=0$, $g$= 0.05, $\lambda$= 1, and $N = 400$.}
\label{BIAOTI}
\end{figure}

\begin{eqnarray}
|F_{\mu\nu}(t)|&=&\mathop{\prod}_{k>0}^M[1-\sin^2(2\Theta_k^{\lambda_\mu})\sin^2(\Lambda_k^{\lambda_\mu}t)
-\sin^2(2\Theta_k^{\lambda_\nu}) \nonumber\\
&&\times \sin^2(\Lambda_k^{\lambda_\nu }t)+2\sin(2\Theta_k^{\lambda_\mu})\sin(2\Theta_k^{\lambda_\nu})\nonumber\\
&&\times \sin(\Lambda_k^{\lambda_\mu}t)\sin(\Lambda_k^{\lambda_\nu}t)
\cos({\Lambda_k^{\lambda_\mu}}t-{\Lambda_k^{\lambda_\nu}}t)-4\nonumber\\
&&\times \sin(2\Theta_k^{\lambda_\mu})\sin(2\Theta_k^{\lambda_\nu})
\sin^2({\Theta_k^{\lambda_\mu}}-{\Theta_k^{\lambda_\nu}})\nonumber\\
&&\times \sin^2(\Lambda_k^{\lambda_\mu}t)\sin^2(\Lambda_k^{\lambda_\nu}t)]^{1/2}.
\end{eqnarray}

When $F_{14}(t)\rightarrow 0$, it denotes that the quantum coherence between the two central qubits is strongly destroyed and undergoes strong decoherence due to  the interaction with the chain environment. When $F_{14}(t)\rightarrow 1$, the two-qubit system is slightly affected by the environment.

\section{\label{sec:level3} CORRELATIONS FOR THE TWO-QUBIT}
 Quantum discord is defined as the difference between the quantum mutual information and the classical correlation shared by two subsystems and expressed as:
\begin{equation}
\mathcal{Q}(\rho_{AB})=\mathcal{I}(\rho_{AB})-\mathcal{C}(\rho_{AB})
\end{equation}
where $\mathcal{I}(\rho_{AB})$ is the total correlation and $\mathcal{C}\left(\rho_{AB}\right)$ represents the classical correlation of state $\rho_{AB}$. These two correlation can be written as:
\begin{figure}
\includegraphics[width=10cm]{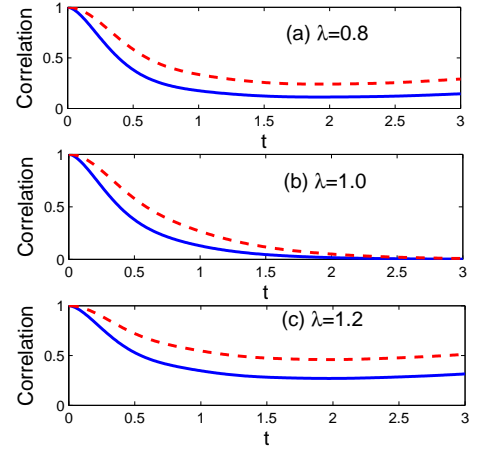}\\
\caption{Quantum discord (solid line) and entanglement(dashed-dotted line) as a function of time t for different $\alpha$, Other parameters
are set as $\gamma$= 1, $\delta=0$, $g$= 0.05, $\alpha$= 0, and $N = 400$.}
\label{BIAOTI}
\end{figure}

\begin{equation}
\mathcal{I}(\rho_{AB})=S(\rho_{A})+S(\rho_{B})-S(\rho_{AB}),
\end{equation}
\begin{equation}
\mathcal{C}(\rho_{AB}) = \max\{S(\rho_{A})-S(\rho_{AB}|\{\Pi_{j}^{B}\})\},
\end{equation}
where $S(\rho)=\mathrm{Tr}\rho\log_{2}$$\rho$ is the von Neumann entropy, $\rho_{A(B)}$ = $\mathrm{Tr}_{B(A)}$($\rho_{AB})$ is the reduced-density operator of partition $A(B)$, $S(\rho_{AB}|\{\Pi_{j}^{B}\})=\Sigma_{j}p_{j}S(\rho_{A|\Pi_{j}^{B}})$ is a quantum extension of the classical conditional entropy, and $\rho_{A|\Pi_{j}^{B}}$ is the reduced state of partition $A$ after the measurement $\Pi_{j}^{B}$ performed on partition $B$, with outcome $j$.

To obtain the quantum discord of $\rho_{AB}(t)$, we need to calculate the quantum mutual information and classical correlation. It is not difficult to calculate the quantum mutual information,
\begin{equation}
\mathcal{I}[\rho_{AB}(t)]=2+\sum_{n=1}^{4}\omega_{n}\log_{2}\omega_{n},
\end{equation}
where $\omega_{1}=\frac{1}{4}(1-c_{3}+|W|)$, $\omega_{2}=\frac{1}{4}(1-c_{3}-|W|)$, $\omega_{3}=\frac{1}{4}(1+c_{3}+|G|)$, and $\omega_{4}=\frac{1}{4}(1+c_{3}-|G|)$. To calculate the classical correlation
$\mathcal{C}(\rho_{AB})$, we take the complete set of von Neumann measurement $\Pi_{B}^{(j)}=|\phi^{(j)}\rangle\langle\phi^{(j)}|$ as a local measurement performed on the subsystem B, where $|\phi^{(1)}\rangle=\cos\theta|0\rangle+e^{i\varphi}\sin\theta|1\rangle$, $|\phi^{(2)}\rangle=e^{-i\varphi}\sin\theta|0\rangle-\cos\theta|1\rangle$ are the two projectors with $\theta,\varphi\in[0,2\pi]$. Thus we obtain the reduced density matrices of subsystem $A$ after measurement,
\begin{equation}
\rho_{A}^{(1)}=\left(\begin{array}{cc}
\frac{1}{2}[1+c_{3}\cos(2\theta)] & \Gamma\\
\Gamma^{*} & \frac{1}{2}[1-c_{3}\cos(2\theta)]
\end{array}\right),
\end{equation}

\begin{equation}
\rho_{A}^{(2)}=\left(\begin{array}{cc}
\frac{1}{2}[1-c_{3}\cos(2\theta)] & -\Gamma\\
-\Gamma^{*} & \frac{1}{2}[1+c_{3}\cos(2\theta)]
\end{array}\right),
\end{equation}
and the probability $p_{1}=p_{2}=\frac{1}{2}$, where $\Gamma=\frac{1}{4}(e^{-i\varphi}W+e^{i\varphi}G)\sin(2\theta)$. Subsequently the classical correlation of Eq. (9) can be calculated as

\begin{figure}
  \includegraphics[width=8cm]{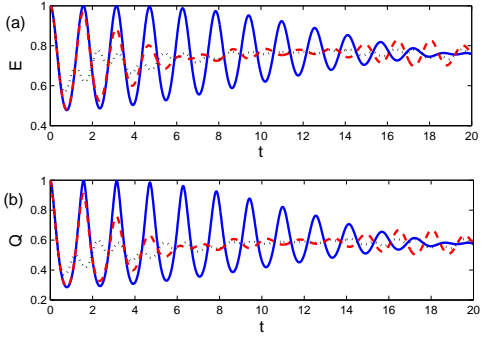}\\

\caption{(a) Entanglement and (b) the quantum discord as a function of time t for different $\alpha$, Other parameters
are set as $\gamma$= 1, $\delta=0$, $g$= 0.05, $\lambda$= 0, and $N = 400$. Here $\alpha=0$ (solid line), $\alpha=0.1$ (dashed line),$\alpha=0.5$ (dotted line)}
\label{BIAOTI}
\end{figure}

\begin{equation}
\mathcal{C}[\rho_{AB}(t)]=\frac{1+\chi}{2}\log_{2}(1+\chi)+\frac{1-\chi}{2}\log_{2}(1-\chi),
\end{equation}
where $\chi=\max\left\{ |c_{3}|,\frac{|W|+|G|}{2}\right\} $. Finally, substituting Eqs. (14) and (17)
into Eq. (11), we can obtain the expression of quantum discord immediately.

In order to compare the dynamical behaviors between quantum discord and entanglement, we need to calculate entanglement. Now we choose entanglement of formation to quantify the amount of entanglement for the two-qubit,
\begin{equation}
E=f\left(\frac{1+\sqrt{1-C^{2}}}{2}\right),
\end{equation}
where $f(x)=-x\log_{2}x-(1-x)\log_{2}(1-x)$ is the binary Shannon entropy, and $C=\max\{0,2\lambda_{max}-{Tr}[\sqrt{\mathrm{\text{\ensuremath{\rho}}_{AB}\varrho}}]\}$ is the
time-dependent concurrence \cite{WKW} with $\varrho=\sigma_{A}^{y}$${\normalcolor \otimes\sigma_{B}^{y}}$$\rho^{\ast}_{AB}\sigma_{A}^{y}{\normalcolor \otimes\sigma_{B}^{y}}$, $\lambda_{max}$ is the the maximum eigenvalue of $\sqrt{\rho_{AB}\varrho}$. We can obtain concurrence for the state given by Eq. (9), $C=\max\left\{ 0,\frac{|G|+c_{3}-1}{2},\frac{|W|-c_{3}-1}{2}\right\}$.

\subsection{\label{sec:level2}Evolution from Pure State}
The dynamical behavior of quantum discord and entanglement for two qubit in XY spin environment have been discussed in \cite{BQLBSJZ}. Here we focus on the dynamic phenomena of the two-qubit due to the three-site interaction. Let us consider the two-qubit with pure initial state. The parameters are set as $c_1=-c_2=c_3=1$, then the initial state becomes Bell state $\frac{1}{\sqrt{2}}(|00\rangle+|11\rangle)$. Substituting these values into Eqs. (11) and (18), we obtain the quantum discord and entanglement of this case, $Q=\frac{1-|F_{14}|}{2}\log_2(1-|F_{14}|)+\frac{1+|F_{14}|}{2}\log_2(1+|F_{14}|)$, $E= 1-\frac{1+\sqrt{1+|F_{14}|^2}}{2}\log_2(1+\sqrt{1-|F_{14}|^2})\
-\frac{1-\sqrt{1+|F_{14}|^2}}{2}\log_2(1-\sqrt{1-|F_{14}|^2})$. Obviously, both of them are monotonically increasing function of variable $|F_{14}\left(t\right)|$ in the interval $[0,1)$. Thus, they behave in a similar way.
\begin{figure}
  \includegraphics[width=10cm]{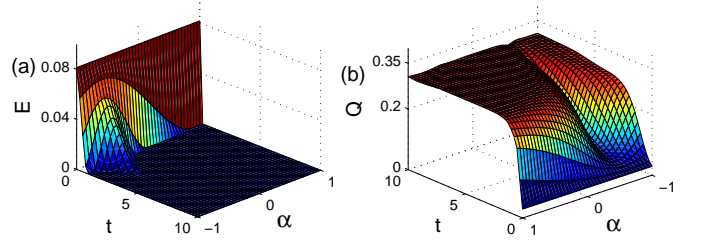}\\

\caption{(a) Entanglement and (b) quantum discord as a function of time $t$ and the three-site interaction $\alpha$, Other parameters
are set as $\gamma$= 1, $\delta$=0, $g$= 0.05, $\lambda$= 1, and $N = 400$.}
\label{BIAOTI}
\end{figure}

Fig. 1 shows the effects of the three-site interaction, which is consistent with the result in \cite{WWCJML}, where the concurrence is discussed in the system. The dynamical behaviors of quantum discord and entanglement are similar as expected and they both decrease monotonously against time. When $t$ is fixed, quantum discord and entanglement behave as parabolas and have maximum value at the same symmetry axis $\alpha=\alpha^\prime$. When $\alpha$ is fixed, quantum discord and entanglement decay to zero in a short time. Furthermore, one can observe that the decay of quantum discord and entanglement become more sharply with increasing intensity of three-site interaction $\alpha$ in the region $\alpha>\alpha^\prime$ and with decreasing $\alpha$ in $\alpha<\alpha^\prime$. At the point $\alpha=\alpha^\prime$, the decay of quantum discord and entanglement be delayed notably. To understand this effect, One may turn to the approximation of $|F_{14}\left(t\right)|$ given in Ref. \cite{WWCJML}. For a weak-coupling parameter $g$ and a large $N$, $|F_{14}\left(t\right)|\thickapprox e^{-\left(\tau_{1}+\tau_{2}+\tau_{3}\right)t^{2}}$ with $\tau_{1}=\frac{32\pi^{2}\gamma^{2}g^{2}}{\left(\lambda-1\right)^{2}}\Sigma(\frac{k}{L})^{2}$, $\tau_{2}=\frac{256\pi^{3}\gamma^{2}g\alpha}{\left(\lambda-1\right)^{2}}\Sigma(\frac{k}{L})^{3}$, and $\tau_{3}=\frac{512\pi^{4}\gamma^{2}\alpha^{2}}{\left(\lambda-1\right)^{2}}\Sigma(\frac{k}{L})^{4}$. When we fix $t$, the exponent $(\tau_{1}+\tau_{2}+\tau_{3})$ is a quadratic function about $\alpha$, so there should be an extremum point where $|F_{14}\left(t\right)|$ reach the maximum value. Since quantum discord and entanglement are monotonically increasing function of variable $|F_{14}\left(t\right)|$ in the interval $[0,1)$, they must reach maximum values at the extremum point. Consequently, the decay of quantum discord and entanglement be delayed at $\alpha^\prime$.

Moreover, from the expression of $|F_{14}\left(t\right)|$, we can see when $\lambda\rightarrow1$, $|F_{14}\left(t\right)|$ will gradually decay against time. Fig. 2 shows the influence of transverse field strength $\lambda$ on the decay behavior of quantum correlations. One can observe that quantum discord and entanglement decrease against time as expected before they tend to stable values, when $\lambda=1$ quantum discord and entanglement suffer sudden death at the same time.
\begin{figure}
  \includegraphics[width=8cm]{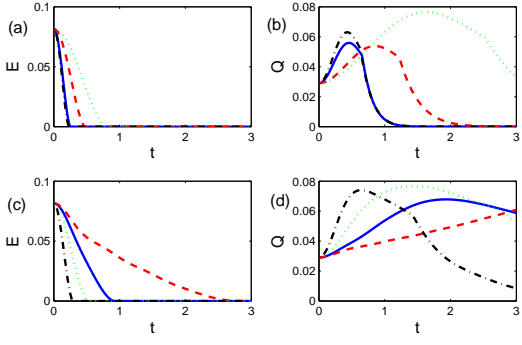}\\

\caption{Quantum discord and entanglement as a function of time t with different $\alpha$, Other parameters
are set as $\lambda$= 1, $\delta$=0.5, $g$= 0.05, $\lambda$= 1, and $N = 400$. Here, $\alpha=-0.8$ (solid line), $\alpha=-0.5$ (dotted line), $\alpha=0$ (dashed line), $\alpha=0.5$ (dashed-dotted line).}
\label{BIAOTI}
\end{figure}

The situation of $\lambda=1$ have been considered in Fig. 1 and we found there exist special interval where the decay of the quantum discord and entanglement can be delayed remarkably.
Here we consider the case of $\lambda=0$. We plot quantum discord and entanglement as a function of time $t$ with different three-site interactions at $\lambda=0$ in Fig. 3.
It is found that the dynamic behaviors of quantum discord and entanglement are obviously distinct from the case of $\lambda=1$.
 The behaviors show oscillatory decay against time. From Fig. 3, one can find the absolute value of amplitude is very sensitive to the three-site interaction. In other words, the absolute value of amplitude is decreasing with increasing three-site interaction $\alpha$, and its decay can be enhanced by $\alpha$. However, the average value of quantum discord and entanglement is approximately remain unchanged. What's more,
the time between the adjacent peak is shorten with increasing the value of $\alpha$. To explain these features, we take a similar tactic employed in Fig. 1. $|F_{14}(t)|=e^{-d\sin^{2}(2\varpi t)}$, where $d=-\frac{8\pi^{2}\gamma^{2}g^{4}}{N^{2}(g-1)^{2}(g+1)^{2}}\sum_{k}^{k_{c}}k^{2}$, $\varpi=1+\frac{4\alpha\pi k}{N}$. Obviously, the exponent of $|F_{14}(t)|$ is quadratic sine function about $\alpha$, and this finding indicate attenuation vibration of the nonclassical correlations, which is consistent with the numerical results shown in Fig. 3.

\subsection{\label{sec:level2}Evolution from Mixed State}
In this subsection, we consider the case of two-qubit initially prepared in the mixed state. We choose the parameters $c_{1}=1$, $-c_{2}=c_{3}=0.2$, so the initial state becomes $\rho_{AB}\left(0\right)=0.6|\phi^{+}\rangle\langle\phi^{+}|+0.4|\varphi^{+}\rangle\langle\varphi^{+}|$, where $|\phi^{+}\rangle=\frac{1}{\sqrt{2}}\left(|00\rangle+|11\rangle\right)$ and $|\varphi^{+}\rangle=\frac{1}{\sqrt{2}}\left(|01\rangle+|10\rangle\right)$.
\begin{figure}
  \includegraphics[width=8cm]{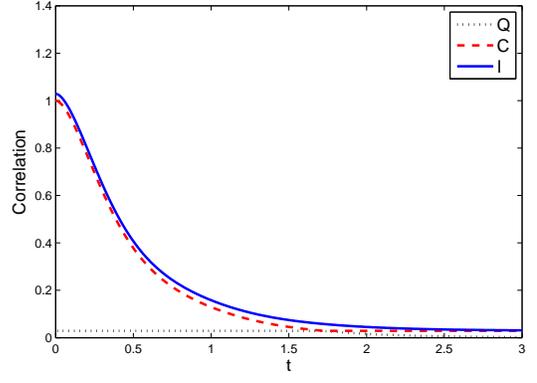}\\

\caption{Quantum discord (solid line), classical correlation (dotted line),
and total correlation (dashed line) as a function of time t. Other parameters
are set as $\alpha=0$, $\gamma$= 1, $\delta$=1, $g$= 0.05, $\lambda$= 1, and $N = 400$. }
\label{BIAOTI}
\end{figure}

In Fig. 4, we show quantum discord and entanglement as a function of $\alpha$ and $t$. One can see that in the special interval nearby $\alpha=-0.5$, the decay of entanglement can be delayed remarkably, while the decay of quantum discord is enhanced. To understand these features, we take a similar tactic employed in Fig. 1. $Q=0.4\log_2(1.6)+0.2+\kappa_{1}\log_2\kappa_{1}-\kappa_{2}\log_2\kappa_{2}$, $E=-\kappa_{3}\log_2\kappa_{3}-(1-\kappa_{3})\log_2(1-\kappa_{3})$, where $\kappa_{1}=0.3+0.3|F_{14}|$, $\kappa_{2}=0.7+0.3|F_{14}|$, $\kappa_{3}=\frac{1+\sqrt{1-[0.6|F_{14}\left(t\right)|-0.4]^2}}{2}$, Obviously, quantum discord is a monotonically decreasing function of variable $|F_{14}\left(t\right)|$, while entanglement is monotonically increasing at interval $|F_{14}\left(t\right)|\in[0.667,1]$ and zero at $|F_{14}\left(t\right)|\in[0,0.667)$. Since there be an extremum point where $|F_{14}\left(t\right)|$ reach the maximum value. So at the extremum point, the decay of quantum discord is enhanced while the entanglement can be delayed.
Moreover, from the plot we find that while the entanglement suffers sudden death after a finite time, the quantum discord is maintained a stable value during the whole time period. This imply that the nonclassical correlations may be more resistant to external perturbations than that of entanglement.

The obvious difference between the behaviors of the quantum discord and entanglement can be found in Fig. 5, where we consider the influence of the anisotropy parameter. One can observe that quantum discord and entanglement behaves in a different way. Quantum discord increase at the first several seconds and then decay with time, while EOF only decay monotonically. At the point of $\alpha=0$, the decay of entanglement can be enhanced by increasing parameter $\gamma$, while the increase of quantum discord is strengthened, and the decay is weaken. For the case of $\alpha\neq0$, we observe that the decay of entanglement can be delayed by increasing the value of $\gamma$. However, under the same situation, both the decay and increase of quantum discord are notably enhanced. What's more, entanglement quickly vanishes in a much shorter time than quantum discord does.
\begin{figure}
  \includegraphics[width=8cm]{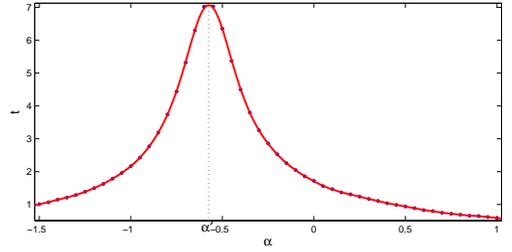}\\

\caption{The transition time as a function of three-site interaction $\alpha$. Other parameters
are set as $\gamma$= 1, $\delta$=1, $g$= 0.05, $\lambda$= 1.}
\label{BIAOTI}
\end{figure}

\begin{figure}
  \includegraphics[width=8cm]{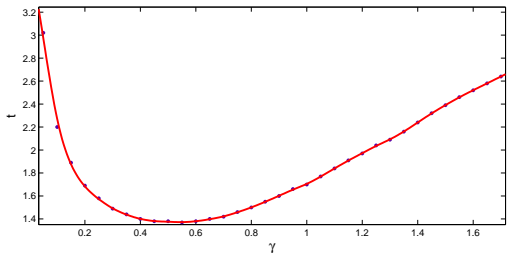}\\

\caption{The transition time as a function of the degree of anisotropy $\gamma$. Other parameters
are set as $\alpha$= 0, $\delta$=1, $g$= 0.05, $\lambda$= 1, and $N = 400$. }
\label{BIAOTI}
\end{figure}

We now turn to investigate the transition behavior of quantum discord which has been discussed in Ref. \cite{MLPJMS}. In Fig. 6, we plot the dynamic behaviors of the quantum discord, the classical correlations and the mutual information against time. We set the parameter $\delta=1$, which means one qubit interacts with the surrounding spin chain and the other is free of the environment. The plot clearly shows the sharp transition from the classical to quantum decoherence and the transition occurs at $t = t'$. In the first $t'$ seconds, the quantum discord is constant and only classical correlation is lost. When $t > t'$, the classical correlation does not change with time and the quantum discord decreases.

Fig. 7 show the transition time is relevant to $\alpha$. One can observe the transition time displays a parabola against time and the symmetry axis locates at $\alpha=\alpha^\prime$. The transition time increase with $\alpha$ in the interval $[1,\alpha^\prime]$ and decrease in $[\alpha^\prime,1]$. The three-site interaction can shorten the transition time $t'$ in most cases compare with the case of $\alpha=0$.
In some special intervals, the interaction can prolong the transition time. This feature can be explained by the analysis employed in Fig. 6. In the special interval of $\alpha$, the decay of quantum discord can be delayed, so it is reasonable to expect quantum discord could keep constant for a longer time than that with other $\alpha$. Namely, the transition time can be prolonged in the special interval. This is important because a stable quantum discord do favor to quantum information processing. A. Brodutch $et$ $al$. \cite{ABDRT} have proposed that the altering of quantum discord is an indicator of failure to a local operations and classical communications in implementation of the quantum gates. So it is Valuable to keep quantum discord a stable value which could be realized via prolonging the transition time.

In Fig. 8, we adopt fitting method to show the dependency relationship between transition time and the degree of anisotropy $\gamma$.
we find $t$ will decrease when $\gamma$ goes from zero to a critical point around $\gamma=0.6$ and increase when $\gamma$ goes from the critical point to the Positive axis direction. At the critical point, the transition time is reach a minimum value. The plot illustrate the transition time can be shorten by the degree of anisotropy $\gamma$.

\section{CONCLUSIONS}
In summary, we have studied the time evolution of quantum discord and entanglement for two-qubit coupled to a spin chain with three-site interaction in the weak-coupling region. We analyzed the quantum discord and entanglement for two-qubit prepared in a class of X structure state. If the system evolute from pure state, we found there exist certain special interval in which the decay of the quantum discord and entanglement can be delayed. At the critical point of the environment, the quantum discord and entanglement suffer sudden death at the same time. If no transverse field is in presence, We found the quantum discord and entanglement displays oscillatory decay with time, the absolute value of amplitude is very sensitive to the three-site interaction. We also considered that two-qubit initially prepared in Bell diagonal state. The decay of entanglement is delayed in the special interval as in the case of pure state, while the decay of quantum discord is enhanced. We have observed sudden transition between classical and quantum decoherence for this case, it is found the transition time can be lengthened at the special interval of three-site interaction and be shorten by degree of anisotropy. Besides, we have considered the effect of anisotropy parameter on the decay behaviors of quantum correlations. It has been shown the effect of three-site interaction on quantum discord can be weaken by the anisotropy parameter.

\section{Acknowledgments}
This work was partially supported by the NSF of China
(Grant No. 11075101), and Shanghai Research Foundation
(Grant No. 07dz22020).

\end{document}